\documentclass[journal]{vgtc}

\onlineid{0}

\vgtccategory{Research}

\vgtcpapertype{please specify}

\title{FINCH: Locally Visualizing Higher-Order Feature Interactions in Black Box Models}

\author{%
  \authororcid{Anna Kleinau}{0000-0002-3415-6316},
  Bernhard Preim, and 
  Monique Meuschke
}

\authorfooter{

  \item all authors are with Otto-von-Guericke University Magdeburg, Germany.
}

\abstract{%

In an era where black-box AI models are integral to decision-making across industries, robust methods for explaining these models are more critical than ever. While these models leverage complex feature interplay for accurate predictions, most explanation methods only assign relevance to individual features. There is a research gap in methods that effectively \emph{illustrate} interactions between features, especially in visualizing higher-order interactions involving multiple features, which challenge conventional representation methods.

To address this challenge in local explanations focused on individual instances, we employ a visual, subset-based approach to reveal relevant feature interactions. Our visual analytics tool \textsc{Finch} uses coloring and highlighting techniques to create intuitive, human-centered visualizations, and provides additional views that enable users to calibrate their trust in the model and explanations. 
We demonstrate \textsc{Finch} in multiple case studies, demonstrating its generalizability, and conducted an extensive human study with machine learning experts to highlight its helpfulness and usability.
With this approach, \textsc{Finch} allows users to visualize feature interactions involving any number of features locally.
}


\teaser{
    \centering
    \includegraphics[width=1\linewidth]{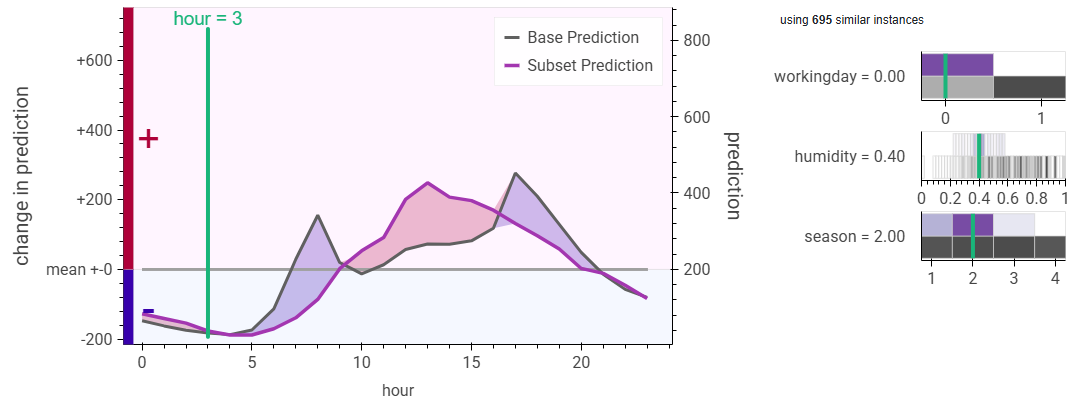}
    \caption{Our visual analytics tool \textsc{Finch} generates dependency plots that can locally visualize feature interactions up to high orders.}
    \label{fig:teaser}
}

\graphicspath{{figs/}{}{pictures/}{images/}{./}} 

\usepackage{tabu}                
\usepackage{booktabs}                 
\usepackage{lipsum}                    
\usepackage{mwe}

\usepackage{mathptmx}                 
\usepackage{dirtytalk}

\begin{document}

\firstsection{Introduction}

\maketitle

As AI grows due to the increasing availability of data, high-dimensional datasets with numerous instances are becoming the norm. Explainable AI (xAI) is evolving alongside this trend; however, traditional xAI methods are reaching their limits as modern machine learning models increasingly rely on interactions among countless features. Explaining features independently or focusing solely on one or two-feature interactions is insufficient. New methods are needed to make higher-order feature interactions —complex interactions among multiple features— understandable for humans, as these often contain the true predictive power of AI models.

Our focus is on local explanations that clarify the prediction for individual instances. For example, in a decision support system, which feature interactions drive the current recommendation? This need becomes evident when considering the following example by Tsang et al.~\cite{tsang2021interpretable}, who describes the prediction function $f(x)=x+y+xy$. If relevance is assigned only to individual features, how do we attribute the interaction between $x$ and $y$? Similar issues arise in practice; Wang et al.~\cite{Wang2019DesigningTU} found that domain experts prefer to see feature interactions, such as the relationship between systolic and diastolic blood pressure. Moreover, from a security perspective, Xin et al.~\cite{xin2024you} show that partial dependence plots focused on a single feature can be manipulated to conceal discriminatory behavior. These examples underscore the growing need for more effective methods to characterize feature interactions.

Our visual analytics tool \textsc{FINCH} was designed to tackle this challenge for high-dimensional, tabular data. Its intuitive user interface allows machine learning experts to easily trace how multiple features interact and see how each additional feature influences the outcome. This is achieved through a subset-based approach using visualizations similar to the well-known partial dependence plots~\cite{friedman2001greedy}. Instead of visualizing all possible interactions in a feature space, our approach focuses on the local interactions relevant to the current data instance. In summary, we make the following contributions: 

\begin{itemize}
    \item We introduce a novel approach for visualizing higher-order feature interactions at the instance level.
    \item We developed an interactive online tool tailored for machine learning experts. 
    \item We conduct a comprehensive evaluation of our tool through both case studies and a user study.
\end{itemize}

\section{Background}
xAI methods are classically divided into \textit{local} and \textit{global} explanation methods~\cite{danilevsky2020survey}. Global methods aim to explain the entire model at once by describing the overall relevance of input features for predictions. Feature interactions are visualized to show how one or two features interact with the prediction. In contrast, local methods focus on explaining the model's behavior for a specific instance, highlighting the relevant features for that instance's prediction and visualizing their influence as negative or positive. This section reviews two of the most frequently used methods.

\subsection{PDPs - Partial Dependence Plots}
\label{sec:pdp}
Partial Dependence Plots (PDPs)~\cite{friedman2001greedy} are a global explanation method that visualizes the influence of one or two features on a prediction, typically using a line chart or heatmap. The x-axis represents the values of a feature, while the y-axis shows the mean prediction. PDPs work well with smaller datasets because they use a mutation-based approach to generate sufficient data points. To calculate the average prediction for a feature value, all available data instances are set to this value, while keeping their other features at their original value.  

A visualization similar to PDPs is the ICE plot. While calculated similarly, each instance is given its own curve instead of averaging all instances to produce a single curve. This results in a more detailed but also more cluttered visualization.

\subsection{SHAP plots}
SHAP plots~\cite{lundberg2018consistent} are a local explanation method based on game theory that visualizes the positive and negative influences of features on the current instance. Each feature is represented by a bar, where the length depicts the strength of its influence, and the direction and color indicate whether the influence is positive or negative. This visualization shows the influence of each feature individually without displaying their interactions, which can be confusing when variables are highly correlated. In such cases, their combined influence might be assigned to one feature or split between them, leading to potentially misleading variations in the final SHAP plot.

Visualizations similar to SHAP, but with different calculation methods, include LIME plots, which are created by learning a model around the local instance~\cite{ribeiro2016should}.

\section{Related Work}
In this section, we introduce the emerging concept of \emph{regional explanations} that inspired our tool, and discuss prior work on interpreting feature interactions and visual analytics in xAI.

\subsection{Regional methods}
Set between local and global methods,
an emerging category of explanation methods are regional methods, which focus on specific subgroups of the data. \textsc{Repid}~\cite{herbinger2022Repid} analyzes the ICE curves of a feature by identifying subgroups with different behaviors based on the shapes of individual instances' curves. A decision tree divides these groups while maintaining interpretable labels for easier understanding.

While \textsc{Repid} requires selecting an initial feature for exploring interactions, \textsc{Gadget}~\cite{herbinger2023decomposing} offers a more general approach by recursively decomposing the entire feature space to identify simple subspaces with minimal interactions, allowing for straightforward models in explanations. Like \textsc{Repid}, \textsc{Gadget} uses decision trees to guide the decomposition process.

Unlike \textsc{Finch}, \textsc{Repid} and \textsc{Gadget} do not identify feature interactions relevant to a specific instance; instead, they search for feature interactions globally by identifying subgroups.

\subsection{Feature Interaction Interpretation}
Few approaches exist for interpreting or visualizing higher-order feature interactions involving more than two features. Tsang et al.~\cite{tsang2021interpretable} provide a comprehensive review of feature interactions, emphasizing the need for improved interactive visualizations, which \textsc{Finch} aims to address. Zhang et al.~\cite{zhang2023capturing} take a mathematical approach, assuming product separability of feature interactions. Their method identifies these interactions and determines the most likely mathematical formula to represent them. Unlike \textsc{Finch}, this approach focuses on global feature interactions and is limited to this specific type of interaction. Friedman~\cite{friedman2024function} visualizes model interactions by breaking them down into a tree structure, allowing for the visualization of up to three features through line or bar charts. This tree provides a global overview of the interactions within the model, whereas \textsc{Finch} focuses on local feature interactions and can visualize interactions of more than three features.

Other approaches group features together, interpreting them as a unified group, which is especially useful for highly correlated features. For instance, Jullum et al.~\cite{jullum2021efficient} proposed computing Shapley values for groups rather than individual features. Ferretini et al.~\cite{ferrettini2022coalitional} also use grouping but as a step to enhance the calculation of final individual feature values. Going further, Mijolla et al.~\cite{mijolla2020human} propose an approach based on latent representations that redefine features as combinations of the original ones, making them easier for humans to understand. An automatic approach to latent representations are dimension reduction techniques like PCA, as demonstrated by Seedorff and Brown~\cite{seedorff2021totalvis}.

\subsection{Visual Analytics for xAI}

Our work aligns with a series of visual analytics tools designed for xAI. \textsc{Prospector}, by Krause et al.~\cite{krause2016interacting}, combines local and global explanation methods with an instance-centric interface. It builds on classical methods like PDPs but lacks functionality for higher-order feature interactions.

\textsc{Vine}, developed by Britton~\cite{britton2019vine}, is specifically designed to visualize feature interactions. Unlike \textsc{Finch}, it follows a global approach, clustering ICE curves for each feature, similar to \textsc{Repid}, and embedding this clustering in an overview and detailed views.

A similar approach was presented by Molnar et al.~\cite{molnar2023model}, who aimed to improve misleading PDPs by finding subgroups in the data with low feature interactions. Like \textsc{Gadget}, they compute a decision tree but visualize the results of each node in a PDP plot of a feature.

Hohman et al.~\cite{hohman2019gamut} designed a visual analytics system that offers various explanations for machine learning experts, although it does not cover higher-order interactions. They note that participants frequently switched between global and local explanations, highlighting the importance of interactivity.

Inglis et al.~\cite{Inglis2022Visualizing} present techniques for exploring pairwise interactions using matrix layouts of PDPs and network visualizations. \textsc{Finch} focuses on the visualization of interactions of even higher orders.

Lundberg et al.~\cite{lundberg2018consistent} created a VA system that uses SHAP force plots to visualize how different features influence the prediction for one instance over time on a clinical example. 
Their system visualizes all features in an independent manner, whilst ours shows their interaction.

\section{A Subset-Based Algorithm for Local Higher-Order Feature Interactions}
This section describes the general idea that \textsc{Finch} is based on, and how it allows scaling up to higher-order feature interactions.

We will use the publicly available bike-sharing dataset~\cite{fanaee2014event} from the UCI machine learning library~\footnote{https://archive.ics.uci.edu/dataset/275/bike+sharing+dataset} as a running example. This dataset contains hourly bike rental data from 2011 and 2012, with input features including hour, weekday, working day, month, and season.

\subsection{A Preservative Approach to Dependence Plots}
\label{methods_pdps}

The typical way to visualize feature interactions is using PDPs, which permutate the instances of an available data set to calculate predictions for each possible value of a feature, as described in Section \ref{sec:pdp}. 

The primary issue is that the permutation process does not consider the conditional distributions of features, ignoring their dependencies. This makes them less reliable for highly correlated features, as altering one feature without adjusting others can create unrealistic instances~\cite{apley2020visualizing}. For example, changing the month in our bike rental dataset without adjusting the season can lead to instances that cannot exist (e.g. July in winter). Addressing this issue is crucial, as such permutation-based methods can be vulnerable to adversarial attacks that conceal discriminatory behavior~\cite{xin2024you}. 

One proposed solution to this problem is the use of the conditional probability distributions. For example,
Apley et al.\cite{apley2020visualizing} proposed ALE plots to account for the conditional probability of other features.
While their method still introduces slight perturbations, these are less pronounced than before. Each original data point generates two new points with slightly higher or lower values.

As dataset sizes have grown significantly over recent decades, we question whether perturbations are still needed to generate additional data points in explainable AI. We propose using only the original data points, avoiding even slight perturbations. This approach preserves all feature distributions and interactions in the dataset that might otherwise be distorted by artificially generated points.  

This method reduces computational costs since it eliminates the need to generate and predict new data points for each interaction being investigated. Instead, we only need to generate predictions once for the original data points.

This approach offers another advantage: even before model training, we can calculate feature interactions directly on the original dataset by simply using the ground truth values instead of the model predictions. This also enables direct comparisons between model predictions and actual values after model training.

In cases of actual data scarcity, modern techniques for generating new points could still provide a viable solution\cite{figueira2022survey}.

The calculated dependency can be visualized similar to PDPs using a line curve with the feature on the x-axis and the outcome (probability) on the y-axis.

\subsection{Using Subsets for Higher-Order Interactions}

To scale our approach up to higher-order interactions, we focus on only those interactions relevant to a specific instance rather than attempting to show all possible interactions.

\begin{figure}
    \centering
    \includegraphics[width=0.8\linewidth]{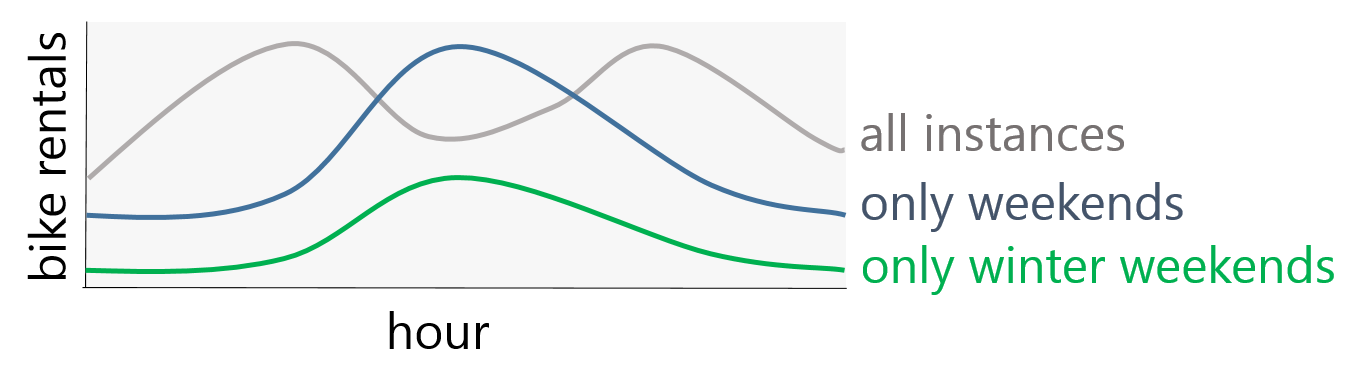}
    \caption{Bike rentals based on different subsets of the data set}
    \label{fig:alg_bikes}
\end{figure}

We want to illustrate our approach on our example of predicting bike rentals (see Fig. \ref{fig:alg_bikes}). When considering only the feature \say{hour}, the predicted dependency curve typically shows peaks around the rush hours.
We now want to consider second-order interactions, using the \say{weekday} as our second feature. As our current instance was recorded on a weekend, we create a subset of instances also recorded on weekends and calculate the new curve based on them. This second curve will lack these peaks due to fewer commuters, showing an afternoon peak instead. 
We can refine this further by considering additional features, such as showing bike rentals by the hour on weekends during winter. By applying such filters to the dataset, we can seamlessly scale up to any number of features, providing a more detailed characterization of how the current instance behaves.

\begin{figure}
    \centering
    \includegraphics[width=0.8\linewidth]{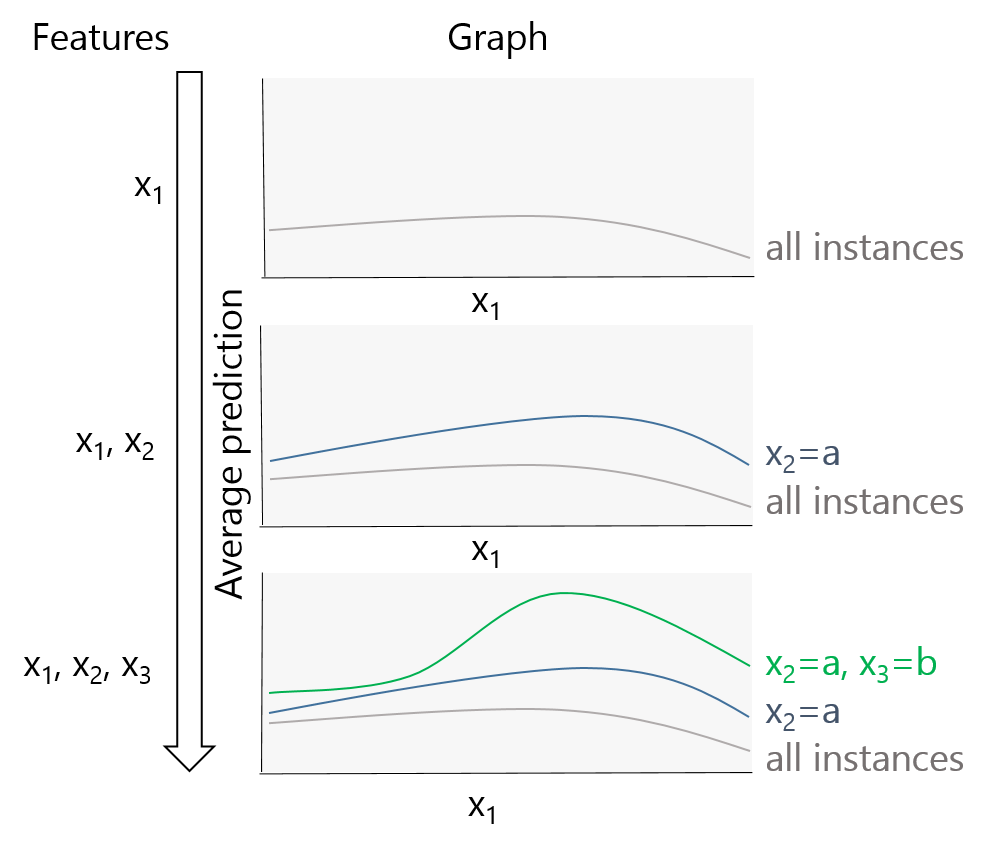}
    \caption{Our algorithm incrementally adding features, by calculating and visualizing subsets for each added feature.}
    \label{fig:algorithm}
\end{figure}

Our algorithm works as follows:
The original line curve is calculated using all instances in the dataset.
For each additional feature, we consider how it interacts with the previous ones when fixed to the current instance's value.
We calculate a second line based solely on instances from the dataset where the second feature matches this value, ensuring we include only realistic, pre-existing data points.
By incrementally calculating a new curve each time a new feature is added, the user can observe how each new feature influences and interacts with previous ones (Fig. \ref{fig:algorithm}).

\subsection{Calculating Subsets}
Ideally, our subsets would contain only instances that have identical feature values as our target instance for the selected features (except for the first one shown on the x-axis). However, since our dataset is limited in size and continuous features rarely have exact matches, we use heuristics to select instances for each subset. These heuristics balance two goals: maintaining high similarity within the subset while ensuring enough data points for reliable calculations.

To achieve this, we employ a set of approximations chosen experimentally:
\begin{itemize}
    \item \textbf{include the 5\,\% most similar instances}
    \item \textbf{include at least the 50 most similar instances.} This makes sure our subset will not be too small and statistically insignificant.
    \item \textbf{include all instances that are almost identical.} this is particularly important for categorical features, where many instances share the same feature value.
\end{itemize}

Similarity is calculated using Euclidean distance between the selected features. Each feature is normalized beforehand to ensure equal weighting.
This approach works for categorical features too, since they must be numerically encoded for use in \textsc{finch}.
By calculating similarity only on selected features, we have access to many more instances compared to requiring similarity across all features.

Instances are treated as almost identical to the current instance if their distance is smaller or equal to the number of columns times 0.1: $d <= len(columns) * 0.1$. 
This value was chosen experimentally.

As our procedure ensures at least 50 instances to be selected for each subset, when an instance does not have highly similar instances, the subset may contain instances that are less similar. Thats why in \textsc{finch}, we caracterize the subset using distribution plots to let the user know how similar the instances are, allowing them to gauge the credibility of the curve.

\subsubsection{Categorical Features}
Our current algorithm adapts to a categorical target by letting the user select one of the classes and visualizing the probability of belonging to that class on the y-axis. 
Further categorical features can be entered into the tool in two ways: through numerical encoding (which creates an implicit ordering) or through one-hot encoding (where each possible feature value becomes its own feature).
When using categorical features, experts should note that numerical encoding creates an implicit ordering that affects similarity calculations between instances. With one-hot encoding, \textsc{finch}'s ability to handle higher-order interactions enables it to consider relationships across all encoded feature values.

\section{Presenting \textsc{FINCH}}
Based on our subset-based algorithm, we designed the visual analytics tool \textsc{Finch} that enables machine lerning experts to incrementally explore feature interactions.
In this section, we outline the initial requirements for \textsc{Finch}, explain its core concept, discuss how the tool facilitates trust calibration, and describe its implementation.

\subsection{Requirements}
The requirements were derived from literature review, as well as discussions with three machine learning experts of varying expertise (ranging from grad student to professor). The discussions were guided by using a rain prediction example, from which they described or sketched the information and visualizations needed to understand feature interactions. They reflected on their past experiences with xAI tools, sharing what they appreciated and what they found lacking. In combination with the literature review, we formulated the following requirements.

\begin{itemize}
    \item (R1) Understandable feature interactions:
    Fundamental to our tool was making feature interactions understandable to experts. Using explanations without completely understanding them often leads to false assumptions~\cite{kaur2020interpreting}.
    \item (R2) Differentiate between positive and negative contributions: 
    Machine learning experts prefer to see how each feature or feature interaction contributes to the prediction, whether positive or negative. This is one of the basic requirements for feature attribution methods outlined by Lundberg et al.~\cite{lundberg2018consistent}.
    \item (R3) Highlight subgroups, outliers, and special characteristics: 
    The experts wanted to know what was unusual about the current instance, which is also seen as the most important information in an explanation in the social sciences~\cite{miller2019explanation}.
\end{itemize}

\subsection{Visual Design of the Dependence Plots}
Our plots are designed to be interpretable in a smiliar manner to PDPs, making it easy for experts to transition between them. The line plots display the relationship between features (x-axis) and predictions (y-axis) through a clear curve. To enhance usability, we incorporate several visualization techniques (Fig. \ref{fig:dependency_plot}).

\begin{itemize}
    \item \textbf{centered around mean}: To enhance differentiation between positive and negative contributions (R2), a horizontal line marks the mean prediction, making it easy to see when predictions are above or below it.
    \item \textbf{colored background}: To further emphasize this, the background is colored in blue below the line, and red above it.
    \item \textbf{+/- symbols}: The two symbols are used to indicate that values the areas indicate predictions above, or below the line.
    \item \textbf{two axes}: Accordingly, the y-axis is also centered around the mean prediction. To allow reading the absolute values, a second y-axis is positioned on the right side of the plot. 
    \item \textbf{highlight whats important}:  We use highlighting through saturated red and blue tones that stand out from the background, depending on the specific visualization.
    \item \textbf{mark current instance}: The x-axis feature value of the current instance is marked by a green vertical line.
\end{itemize}
 
\subsection{Visualizing Interactions Incrementally}

\begin{figure}[h!]
    \centering
    \begin{subfigure}[b]{0.5\textwidth}
        \centering
        \includegraphics[width=1\linewidth]{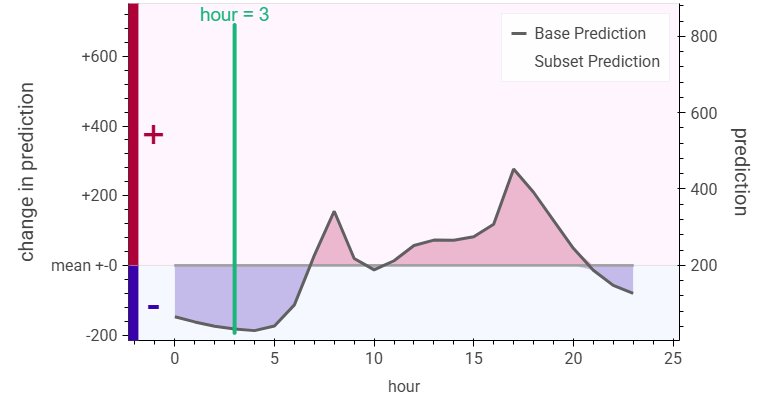}
        \caption{bike rentals per hour}
        \label{fig:dependency_plot}
    \end{subfigure}
    \hspace{0.001\textwidth}
    \begin{subfigure}[b]{0.5\textwidth}
        \centering
        \includegraphics[width=1\linewidth]{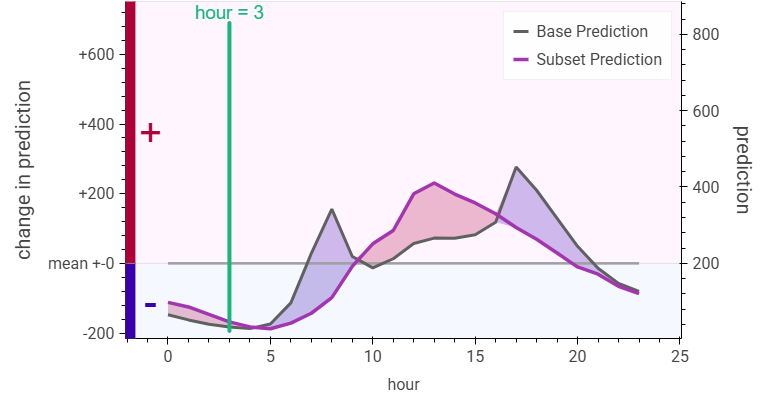}
        \caption{bike rentals per hour on weekends}
        \label{fig:weekend}
    \end{subfigure}
    \hspace{0.001\textwidth}
    \begin{subfigure}[b]{0.5\textwidth}
        \centering
        \includegraphics[width=1\linewidth]{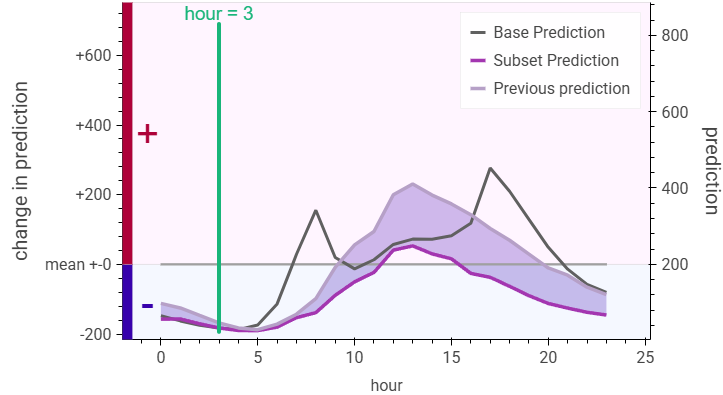}
        \caption{bike rentals per hour on weekends in winter}
        \label{fig:winter_weekend}
    \end{subfigure}
    \caption{The incremental visualization of the interaction of hour, weekends, and winter in the bike rental data set. Colored areas visualize the change in each step.}
\end{figure}

The crucial point of our tool was combining the individual curves for each incremental subset into an intuitive explanation.
For each additional feature added to the current interaction, a new curve is generated.
Simply showing all curves at once quickly leads to cluttered visualizations
We therefore carefully selected which curves to include at each step of the process.
The base curve, calculated based on all instances, is included in all visualizations. This ensures a consistent base for comparison. It is displayed in an unobstrusive grey.
The current subset is displayed as a purple curve, being the most prominent curve of the plot.
The previous subset is visible as a desaturated purple curve to enable comparison.
All older curves are hidden to reduce visual clutter.

We employ highlighting techniques to lead the experts attention to the change (Fig. \ref{fig:dependency_plot} - \ref{fig:winter_weekend}).
\begin{itemize}
    \item \textbf{one feature}: Only the base prediction is visible, showing the base prediction across the values of the first feature, that is depicted on the x-axis. We highlight the difference of the base prediction to the mean (Fig. \ref{fig:dependency_plot}).
    \item \textbf{two features}: A purple line is added for the first subset generated from the new feature. We highlight the difference between the base prediction and the subset (Fig. \ref{fig:weekend}).
    \item \textbf{three or more features}: The previous subset is shown as a desaturated purple line, and the new subset displayed in purple. We highlight the difference between the last subset and the current subset (Fig. \ref{fig:winter_weekend}). Alternatively, the expert can choose to highlight the difference of the current subset to the base prediction instead.
\end{itemize}

\subsubsection{Separating Main and Interaction Effects}
\label{sec:interaction}
One of the optional visualizations that \textsc{Finch} provides is the separation of main and interaction effects.

When adding a new feature to the subset, we highlight the difference between the previous prediction without, and the current prediction including it. This difference can be divided into two components: the main effect of the newly added feature and its interaction effect with previously included features.

In the bike rental scenario, in winter there are overall fewer bike rentals due to its main effect. Additionally, winter interacts with hour, resulting in even fewer bike rentals in the morning but slightly more in the evening on winter days.

Mathematically, we can summarize the previously predicted curve as $c + f_X(y)$ where  $c$ is the mean prediction (as our visualizations center around the mean) and $f_X(y)$ the influence of the previous subset $X$ on the variable on the x-axis, $y$. When we add a new feature $Z$, the updated function becomes $f_{X,Z}(y) = c + f_X(y) + a_Z + g^X_Z(y)$, where $a_Z$ represents the main effect of $Z$, and $g^X_Z(y)$ captures the interaction effect of $Z$ with all previous features. 
We can calculate $a_Z$ independently of all other features as the average of all instances that are in the subset defined by only $Z$.   

\begin{figure}
    \centering
    \includegraphics[width=1\linewidth]{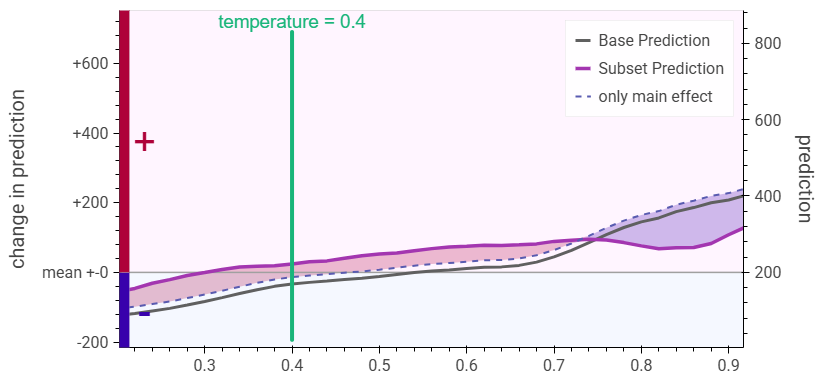}
    \caption{Interaction effect visualization. This visualization separates the main and interaction effects of a newly added feature by showing the main effect through a blue dotted line and the interaction effect through its difference from the actual purple line.}
    \label{fig:interaction_effect}
\end{figure}

To help experts distinguish between these effects, we provide an option to display a new line in our plot showing only the main effect of the newly added feature added to the previous prediction. Since the main effect is independent of all other features, including the one on the x-axis, it shifts the previous curve up or down by a constant. We then highlight the difference between this new line and the actual prediction, revealing the interaction effect between the new feature and previously added features (Fig. \ref{fig:interaction_effect}).

\subsubsection{Choosing the Next Feature}
We use small multiples to help experts choose the next feature that should be added to the currently visualized interaction.

For each potential feature to be added to the subset, we display its dependence plot, ranking features by the strength of their interaction effect at the instance's x-axis value. 
The user can choose to either view the standard dependence plots, or those displaying the interaction effects ass described in Section \ref{sec:interaction}.

\subsection{Calibrating Trust}
xAI is subject to various cognitive biases~\cite{bertrand2022cognitive}. For instance, merely having an explanation, even if not meaningful, can increase trust in the explained prediction~\cite{eiband2019impact}. Our explanations are approximations of the model's behavior based on the provided datasets, and the model may not always accurately reflect the data. Thus, it is crucial to provide machine learning experts with a variety of tools to calibrate their trust in both the explanations and the model to combat cognitive biases.

Displaying additional information directly on the existing plots was deemed disadvantageous, as it made our plots too cluttered. We therefore decided to display the information in separate plots, or provide different views of the plots that hide aspects of it to make space for the new information. We aimed to make each plot version look slightly different, to help experts in keeping track of which view they are currently seeing, whilst still keeping visual consistency to enable easy interpretation (R1).

\subsubsection{Subset Characterization}
First, we provide a detailed characterization of the subset used for our calculations, including the number of instances and their distribution across selected features compared to the general distribution. The distributions of features used for neighborhood calculations are shown on the right, while the distribution for the x-axis feature is displayed directly below the dependence plot for quick reference.

\begin{figure}
    \centering
    \includegraphics[width=0.7\linewidth]{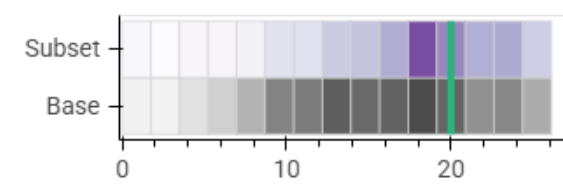}
    \caption{Distribution heatmap. Directly below the dependence plot are the density distributions for the first selected feature, which is visualized on the x-axis.}
    \label{fig:density}
\end{figure}

Both distributions are visualized using heatmaps (Fig. \ref{fig:density}). Their brightness reflects the relative number of instances per distribution rather than the absolute count. This ensures that the distribution of the subset is visible, even when it is much smaller than the data set. The current instance is indicated by a green line, as in the dependence plot.

\subsubsection{Ground Truth}
\begin{figure}
    \centering
    \includegraphics[width=1\linewidth]{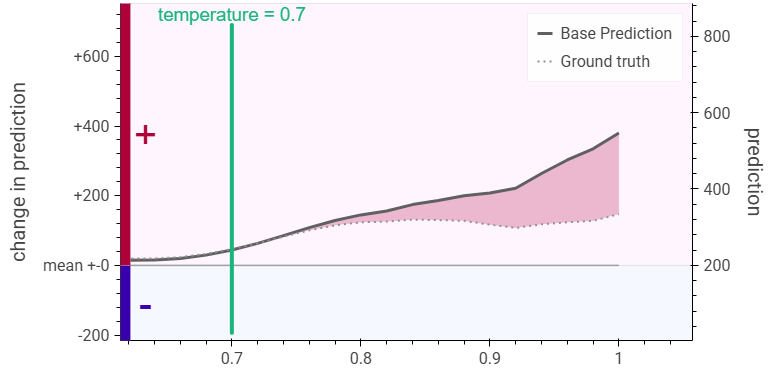}
    \caption{Truth visualization. The deviation between the model prediction and ground truth is highlighted, and the ground truth is displayed through a dotted line.}
    \label{fig:truth}
\end{figure}

\textsc{Finch} uses only real instances, without permuting or generating new ones. By working with actual data, we allow users to provide the ground truth for these instances, enabling direct comparison between the model’s predictions and true values. When the expert selects the ground truth view, it is displayed as a dotted line, with differences between the model's prediction and the ground truth highlighted in blue or red areas. For example, in Figure~\ref{fig:truth}, the model's predictions generally align with the ground truth but exceed it beyond a certain threshold.

\subsubsection{Uncertainty}
The uncertainty induced by mean approximation, as well as the uncertainty innate to the data are addressed through an uncertainty visualization.
This visualization uses a simple area plot to display the standard deviation around the mean for each feature value along the x-axis.

\begin{figure}
    \centering
    \includegraphics[width=1\linewidth]{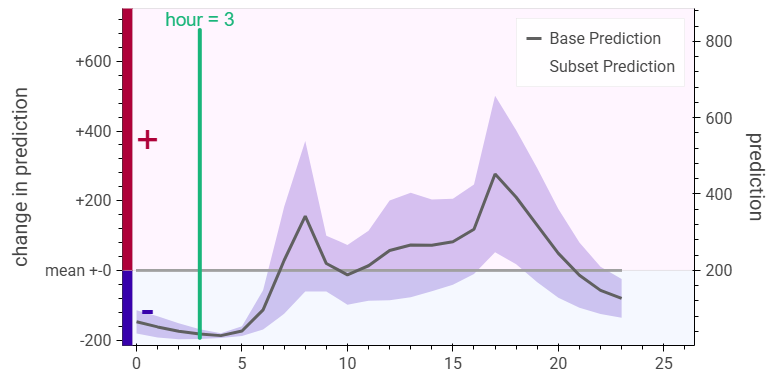}
    \caption{Uncertainty visualization. The standard deviation at each feature value is shown as an area around the curve.}
    \label{fig:uncertainty}
\end{figure}

As shown in Figure~\ref{fig:uncertainty}, the predicted curve shows greater deviation at the peaks and less deviation at lower feature values. It is important to note that this visualization represents the uncertainty of the mean predictions for each feature value, not the model's uncertainty when predicting an individual instance.

\subsection{Implementation}
\begin{figure}
    \centering
    \includegraphics[width=1\linewidth]{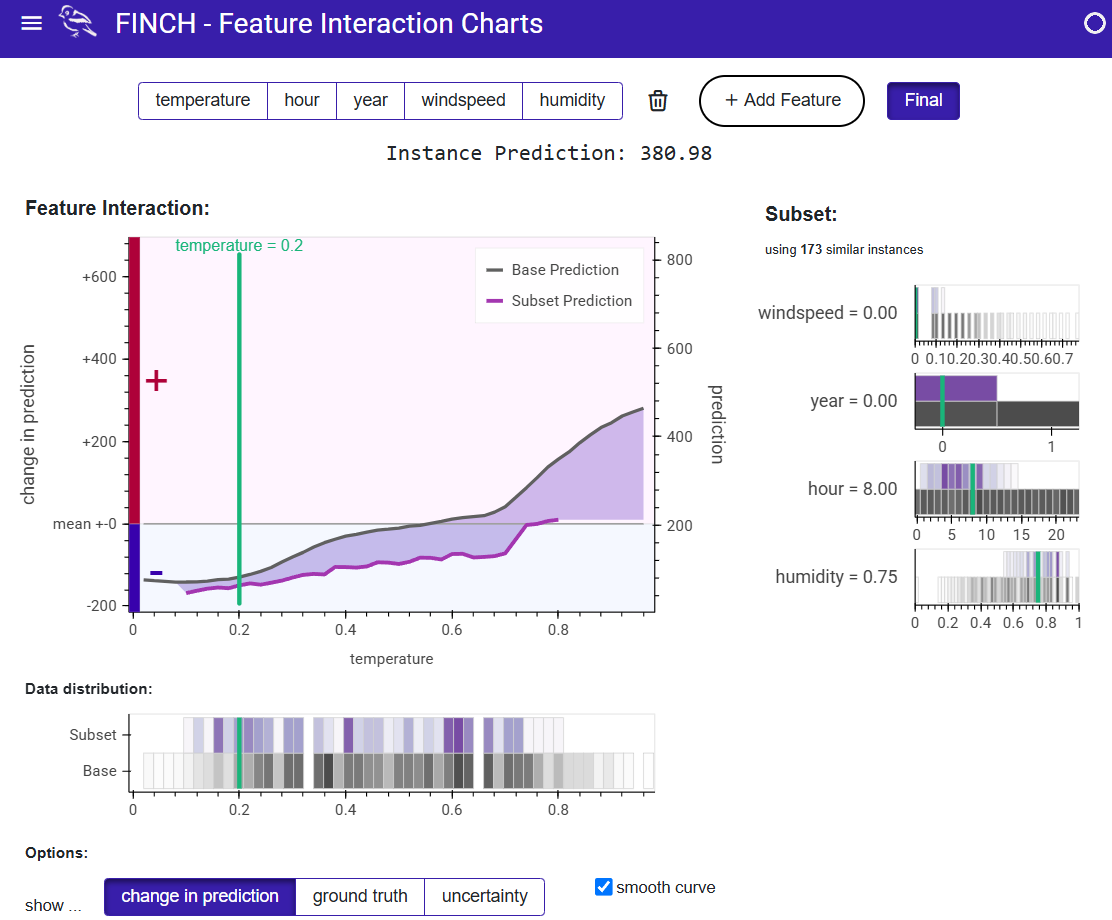}
    \caption{The user interface of \textsc{Finch}. On the left side, the dataset, target variable, and instance can be selected. The main view shows the dependence plot and the distribution heatmaps. On top are options to select features, and on the bottom are different views of the plot.}
    \label{fig:UI}
\end{figure}

\textsc{Finch} is implemented using the Python library Panel\footnote{https://panel.holoviz.org/index.html} for the user interface and Bokeh\footnote{https://docs.bokeh.org/en/latest/docs/gallery.html} for visualizations. The UI of \textsc{Finch} is shown in Figure~\ref{fig:UI}. A test version is hosted on Hugging Face Spaces. Currently, it supports loading all sklearn models saved with python version 3.11 and sklearn version 1.5.1, and datasets saved as CSV files. The code is open source and available on GitHub\footnote{https://github.com/akleinau/Finch}.

The user initially loads a data set (here bike-rentals) and gets an overview of the dependence plots for all features, centered on the current instance's values. Selecting a feature takes the user to the main \textsc{Finch} view, where additional features can be added to refine the subset.

Alternatively, users can load a different data set and model, with the option to provide the ground truth. If there are multiple outcome variables, the desired one can be selected and renamed if needed. Users can choose from the instances of the data set or create a custom one, and any missing values are automatically imputed with the mean.

\textsc{Finch} is tested for both regression and classification models. 
Supported features include continuous and categorical/ordinal types, with any feature having fewer than 24 unique values classified as categorical/ordinal.
Curves of features with more than 24 different values are smoothed using an exponentially weighted average for a smooth appearance that reduces the noise of random statistical fluctuations. The degree of smoothing increases when less data is available or when subsets contain fewer instances. Smoothing can be optionally deactivated.
Categorical feature curves are not smoothed to keep predictions for each class accurate, with the threshold of 24 chosen to preserve the ordinal nature of hourly data.

\section{Case Studies}
We want to demonstrate the usefulness and generalizability of \textsc{Finch} using a set of different data sets with mixed features, different machine learning models, as well as comparisons to common xAI methods. The chosen data sets are inspired by the case studies of Herbinger et al.~\cite{herbinger2022Repid}. Each data set and model is described in detail in Appendix \ref{AppendixA}.

\subsection{Titanic data set}
The titanic data set is used for predictions of the survival of people on board of the titanic, by using features like age, sex and class~\cite{dawson1995unusual}.

\subsubsection{Zero Interactions - Comparison with SHAP}
\begin{figure}
    \centering
    \includegraphics[width=1\linewidth]{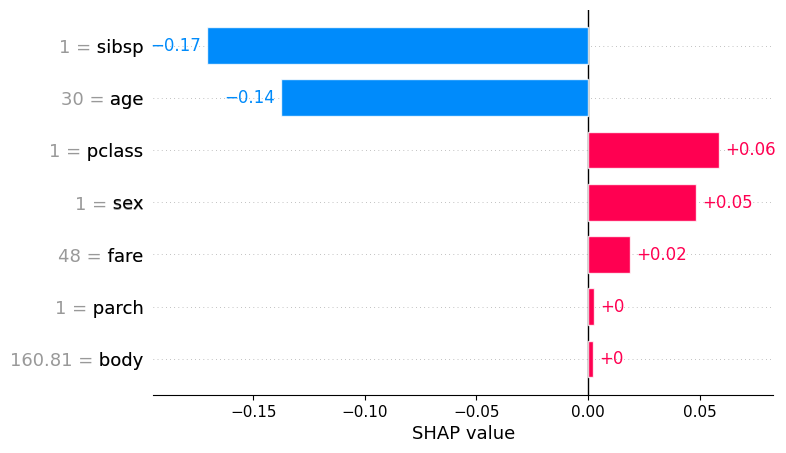}
    \caption{A SHAP plot generated for an instance of the titanic data set. The influence of all features is visualized independently by calculating a \textit{shap value} for each feature, stating its positive or negative influence on the prediciton. Here, the survival of a person on board the titanic is predicted.}
    \label{fig:shap_titanic}
\end{figure}

When analyzing how a prediction for a specific instance came to be, a typical first step would be to generate a \textsc{SHAP} plot showing each features influence. Here, we show the survival chance of a 30yo woman. \textsc{SHAP} gives you a general overview of all features, but also requires all features to be specified (Fig. \ref{fig:shap_titanic}). Feature interactions are considered during the calculation, but they are not visible to the end user. Therefore, the machine learning expert may now ask how specifically class, sex, and age played together. Our tool shows how they interact.

\subsubsection{One or Two Interactions - Comparison with PDP}
\begin{figure}
    \centering
    \includegraphics[width=1\linewidth]{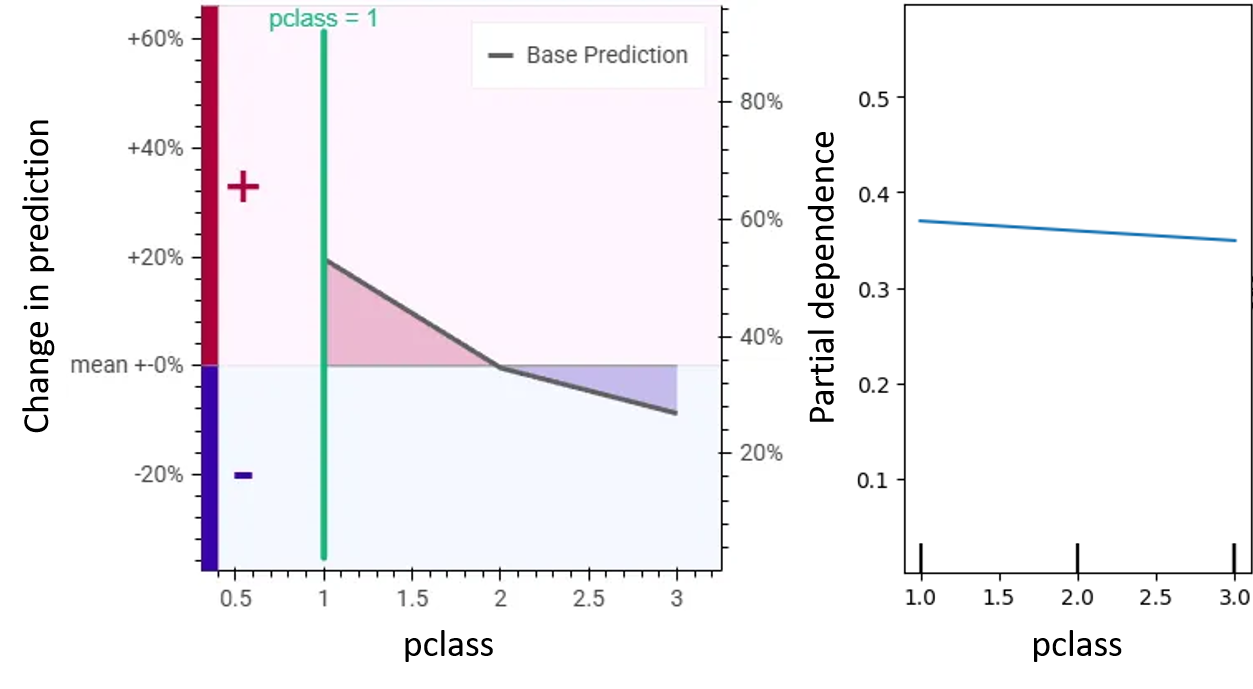}
    \caption{Comparing plots for first order feature interactions. Left: our plot. Right: PDP line chart.}
    \label{fig:pdp_titanic}
\end{figure}

When only considering the interactions of one feature, like class, \textsc{finch} plots look relatively similar to PDPs.
Both show the survival chance per class of the titanic, with first class travelers having the best chance at survival.
However, compared to PDPs, our plots are enhanced with guiding colors and highlights.
The PDP does not give a lot of guidance (Fig. \ref{fig:pdp_titanic}).
What differentiates both plots most is the underlying calculation of probabilities, as described in Section \ref{methods_pdps}. 
The way of PDP generating new data points may result in unrealistic data points, when, for example, additionally to the class, a wealth feature would be available that correlates strongly with it but is not changed accordingly.
In our case, the dependence seen in the PDP is more linear than the one our algorithm has found. 
One explanation for that is by just using realistic data points, our plot takes into account that people who travel in the first class probably also are privileged in other aspects that may lead to heightened survival chances. 

\begin{figure}
    \centering
    \includegraphics[width=1\linewidth]{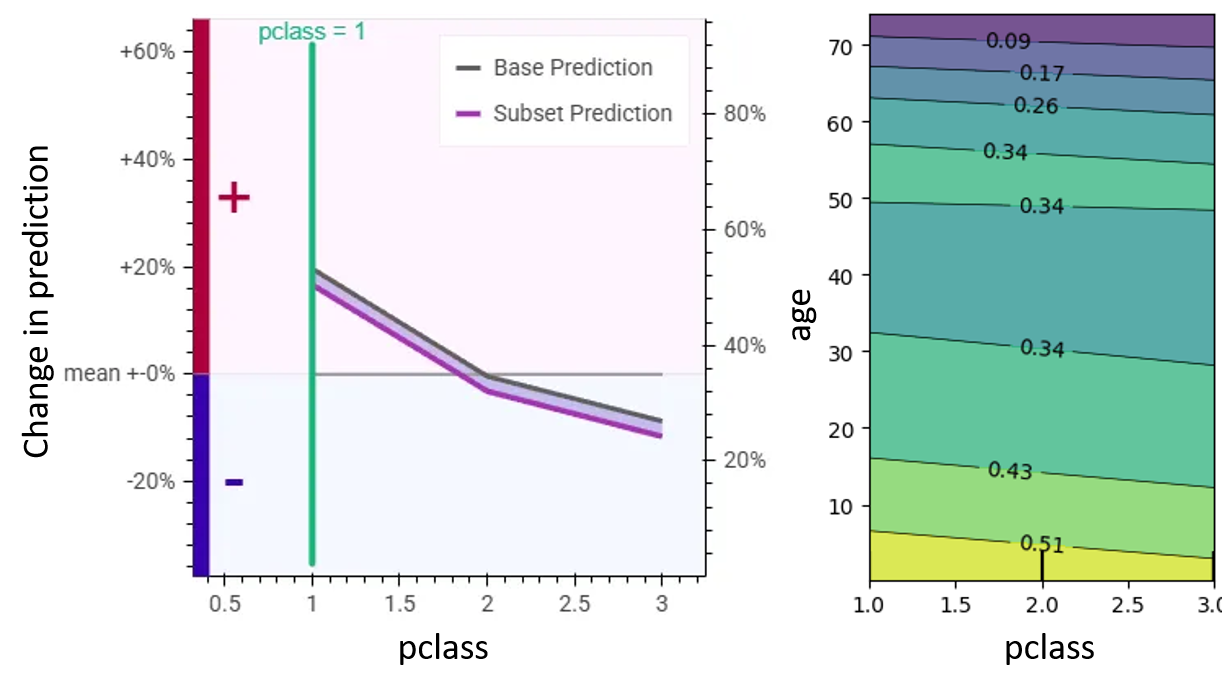}
    \caption{Comparing plots for second order feature interactions. Left: our plot. Right: PDP heatmap. }
    \label{fig:titanic_2F}
\end{figure}

When scaling up to two features by also considering the age, the \textsc{FINCH} plot stays relatively similar, simply adding another line showing the probability change for the current instance of age 30 (Fig. \ref{fig:titanic_2F}).
However, as PDP plots are designed to show global interactions, they are now displayed as heatmaps. This reveals more information, at the cost of requiring significantly more time to interpret.
Additionally, our plot highlights the mean probability, which is not highlighted in the PDP.

\subsubsection{Higher Order Interactions}
\begin{figure}
    \centering
    \includegraphics[width=1\linewidth]{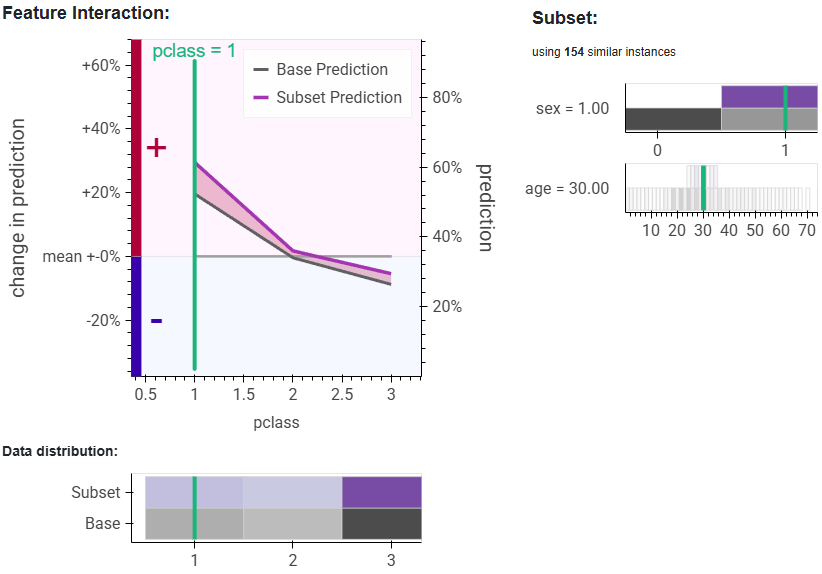}
    \caption{Survival change per class. General public compared to a 30yo woman. Their chance of survival was greater, but especially when in first class.}
    \label{fig:titanic_3F}
\end{figure}

\textsc{Finch} is uniquely able to display feature interactions of more than two features. When additionally considering the gender of the person, the plot shows that 30yo women have a significantly increased probability of surviving, but especially when in first class (Fig. \ref{fig:titanic_3F}). Additionally, as \textsc{finch} is uniquely able to compare this to the ground truth, the dotted line and blue highlighted area display how the model categorically underestimated the survival of those women.

To show the generalizability of \textsc{finch}, we will now display a range of interactions found in other data sets.

As a nonlinear and continuous example, we use the California Housing dataset~\cite{pace1997sparse}. It contains a continuous target variable and continuous features, and is used to predict house values. We trained a gradient boosting regressor model on this dataset.

Using \textsc{Finch}, we successfully identified several higher-order interactions. Detailed depictions of each interaction can be found in Appendix \ref{AppendixA}.
\begin{itemize}
    \item Median income strongly predicts house values, increasing linearly from 2 to 9 and remaining relatively stable above and below these values.
    \item For populations of 2,000, housing values per income behave similarly, except that they stabilize at a lower value for incomes above 8.
    \item However, when these populations have approximately 1,000 total rooms, housing values decrease significantly. For incomes below 4, values remain relatively stable, but then stabilize significantly lower instead of increasing linearly.
    \item A comparison with ground truth for this subset shows the model is relatively accurate, though it slightly overestimates housing values for incomes above 4.
\end{itemize}

The last example features a categorical target and categorical features using a subset of the BRFSS dataset~\cite{BRFSS2015}, which focuses on diabetes risk factors. We trained a decision tree classifier on this dataset.
Using \textsc{Finch}, we observed the following interactions.
\begin{itemize}
\item Diabetes risk in the general population is roughly equal between men and women.
\item Exercise reduces diabetes risk more significantly in women than in men.
\item When combined with high blood pressure, exercise's protective effect is neutralized, resulting in an elevated but equal risk for both men and women.
\item Among those who are also obese, the risk increases further, particularly for women.
\end{itemize}

\section{Evaluation}
Throughout the development of \textsc{Finch}, we continuously tested the tool with various visualization and machine learning experts, leading to significant improvements. To further validate the final version, we conducted a user study with machine learning experts of varying expertise in xAI methods.
We specifically focused on the goals of understandability, usability and helpfulness.

\subsection{Study design}
Participants were recruited by contacting machine learning experts.
The five participants (2F, 3M) had an average age of 35, ranging from 32 to 45.
They work as professors, researchers, graduate and postgraduate students in the field or had significant experience from a previous job.
One participant also participated in the requirement analysis study. None are authors of this paper.
The study was conducted either in person or via online video calls. We began by collecting demographic information and learning about the participant's previous experiences. 
Next, we introduced our bike rental example dataset to create a simple tutorial scenario, providing a brief description of the prediction task followed by a SHAP graph to overview the current instance and its prediction. We chose SHAP as a widely known explanation method~\cite{holzinger2022explainable}. \textsc{Finch} complements SHAP by offering additional functionality for exploring feature relationships. We guided participants through the tool's features using a set of tasks, explaining how it works along the way.

Participants then worked independently with a dataset on diabetes risk factors derived from the BRFSS telephone study~\cite{BRFSS2015}. They were tasked with answering questions to assess their ability to use the tool independently.

Throughout the process, participants were encouraged to verbalize their thoughts using the think-aloud method. At the end of the session, they rated their experience through a series of questionnaires.
We used the widely recognized System Usability Scale (SUS)~\cite{Brooke1996SUS} to evaluate the tool's usability. To assess the provided explanations regarding their helpfulness, we employed the explanation satisfaction scale (ESC) by Hoffman et al.~\cite{hoffman2018metrics}. We also included specific questions to measure how well our tool met requirements and participants' satisfaction with the visualizations (Fig. \ref{fig:Custom1} and \ref{fig:Custom2}). 

\subsection{Study Results}
All participants reported high experience in machine learning. Their familiarity with explainable AI varied from moderate to very experienced. All were well-acquainted with SHAP and PDP, except for one participant being unfamiliar with PDP.

We split up our findings into the three design goals of understandability, usability and helpfulness.

\subsubsection{Understandability}

During the study, all participants quickly grasped the concept of subsets for one feature. One participant described it as \say{nice and intuitive}. Some participants took longer to understand how additional features were visualized when added to the subset, but with some help all understood it and were able to use and interpret it (meet R1). 

While all participants understood the uncertainty visualization, there was initial confusion due to the varying meanings of "uncertainty"; for instance, one participant thought it referred to the model's prediction uncertainty.
The most challenging aspect for participants was the interaction effect visualization. All struggled with it, though three eventually felt they understood it.

\begin{figure}[h]
    \centering
    \includegraphics[width=1\linewidth]{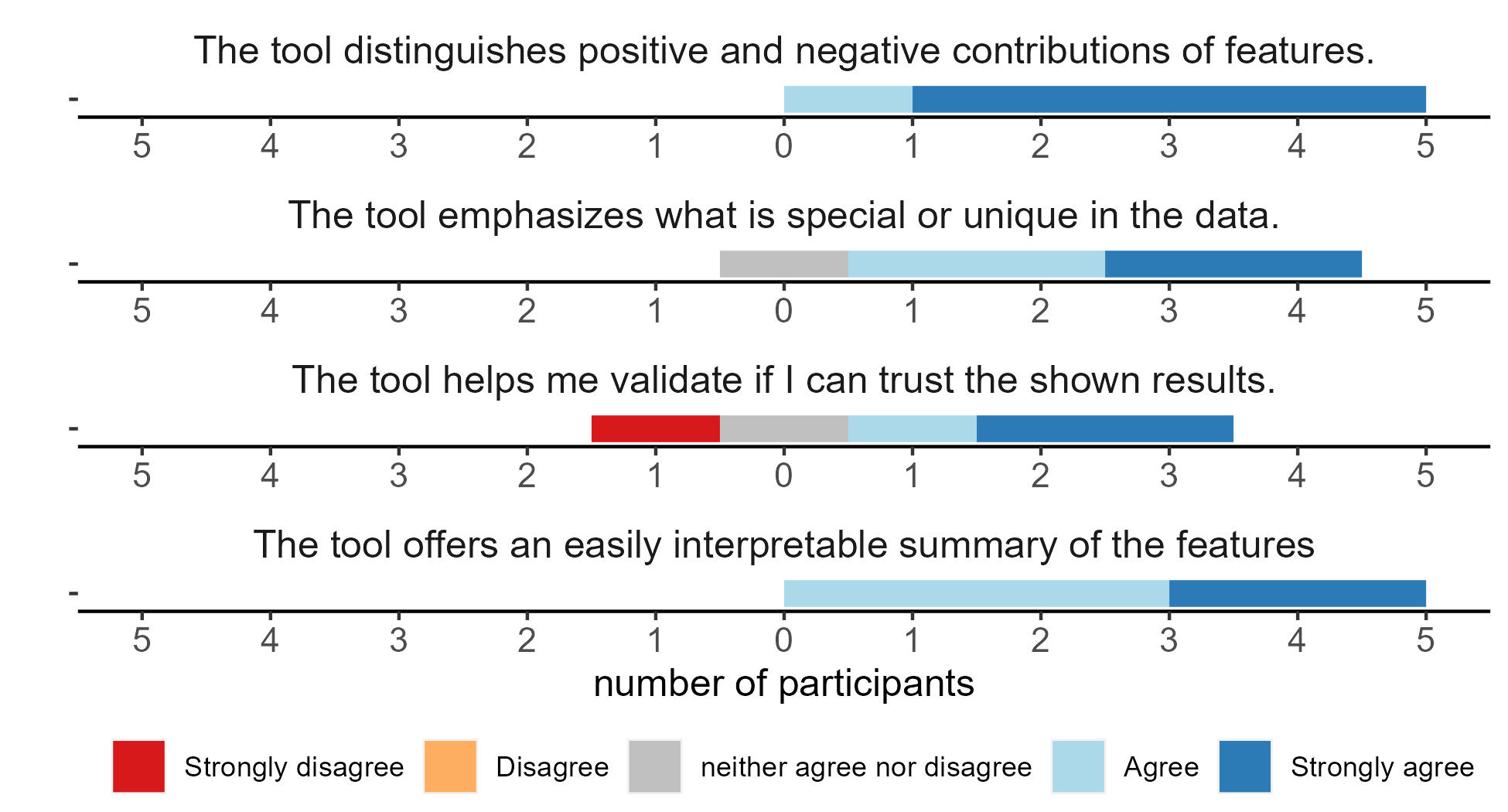}
    \caption{Answers to our custom questions about the tools functionality. For each question, colored bars display the number of people per answer, centered around the neutral answer with disagreeing answers to the left and agreeing answers to the right.}
    \label{fig:Custom1}
\end{figure}

The results from our custom questions were largely positive (see Figure~\ref{fig:Custom1}). Participants agreed that the tool emphasizes unique data aspects (meet R3), distinguishes positive and negative contributions of features (meet R2), and provides an easily interpretable summary of features.
The only negative feedback concerned whether the tool helps validate displayed results. Discussions revealed varied interpretations of this question; while we aimed to assess trust calibration features, some participants interpreted it as evaluating model validation, explaining the mixed responses. 

\subsubsection{Usability}
When using the tool on the diabetes dataset, all participants frequently adjusted feature values to see how predictions changed. Three participants used the ground truth and uncertainty features to validate their results.

One participant suggested increasing contrast for clarity and using textures for colorblind users.
One recommended including a logical description of the current subset for better clarity and a more precise legend for easier recall.
Participants expressed a desire for units to be displayed alongside the numbers. One requested additional information, such as a global explanations, while another wished to move or delete individual features. One participant also wanted to read specific values directly from the chart.
One person raised concerns about scalability, as the tool occasionally ran slowly,
Additionally, one participant found the feature overview overwhelming.
The additional visualizations displaying data distribution, trust, and uncertainty were well received. One participant suggested adding a color legend for the data distribution heatmap. The ground truth visualization was considered very important by one participant, while another expressed interest in using the tool directly on the ground truth for insights before training a model.

\begin{figure}[h]
    \centering
    \includegraphics[width=1\linewidth]{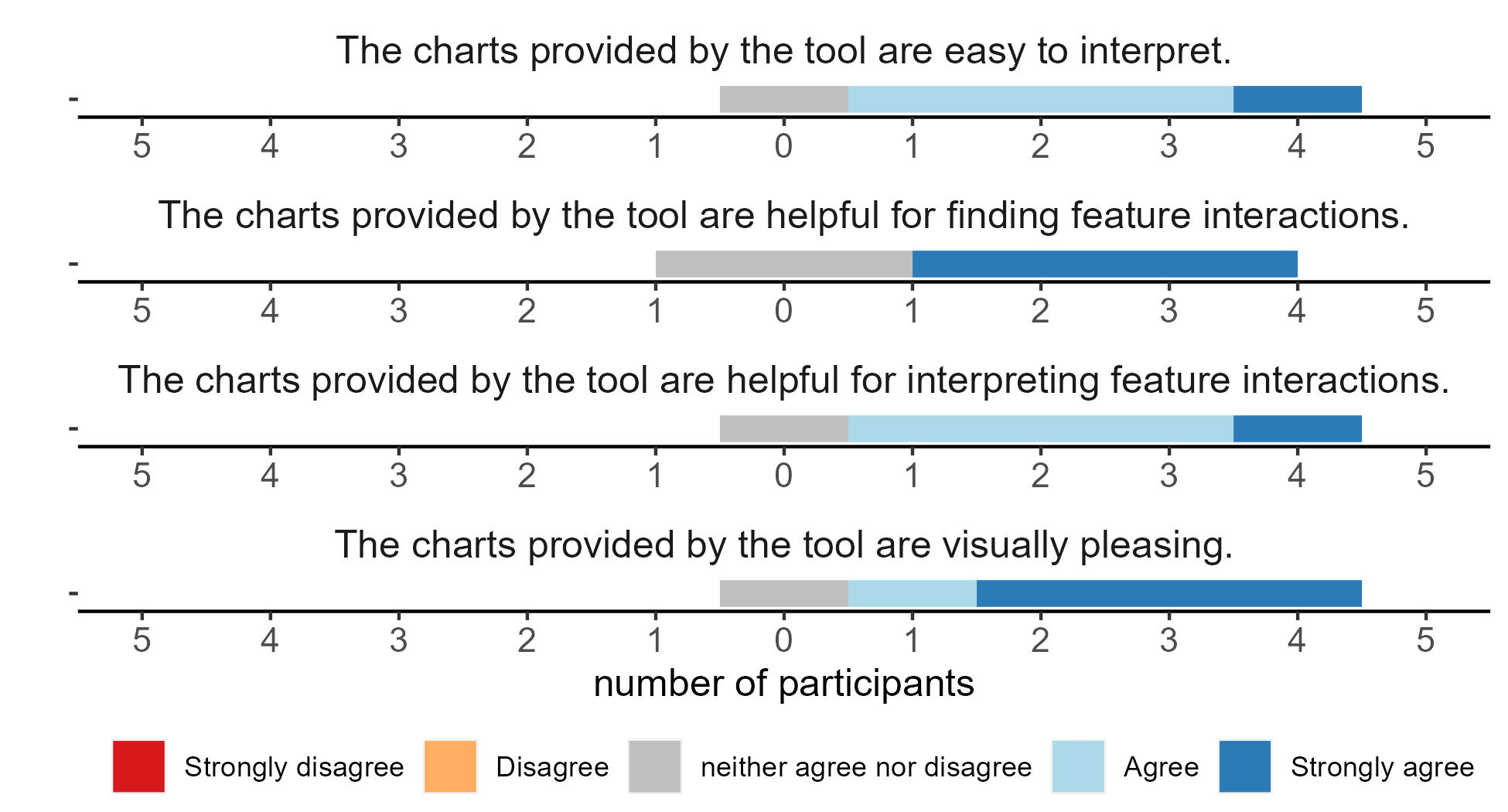}
    \caption{Answers to our custom questions about the visual chart design. For each question, colored bars display the number of people per answer, centered around the neutral answer with disagreeing answers to the left and agreeing answers to the right.}
    \label{fig:Custom2}
\end{figure}

Our tool received a score of 82 on the SUS, indicating excellent usability.
Overall, participants found the charts easy to interpret, visually pleasing, and helpful for identifying and interpreting feature interactions (Fig. ~\ref{fig:Custom2}).
Three experts provided freeform feedback. Two suggested improvements, including, in addition to what was discussed already, an info button to explain the visible interactions and functionality enabling users to \say{play} with the tool by clicking on data points to update values. Two freeform comments also included praise of the tool.

\subsubsection{Helpfulness}
\begin{figure}[h]
    \centering
    \includegraphics[width=1\linewidth]{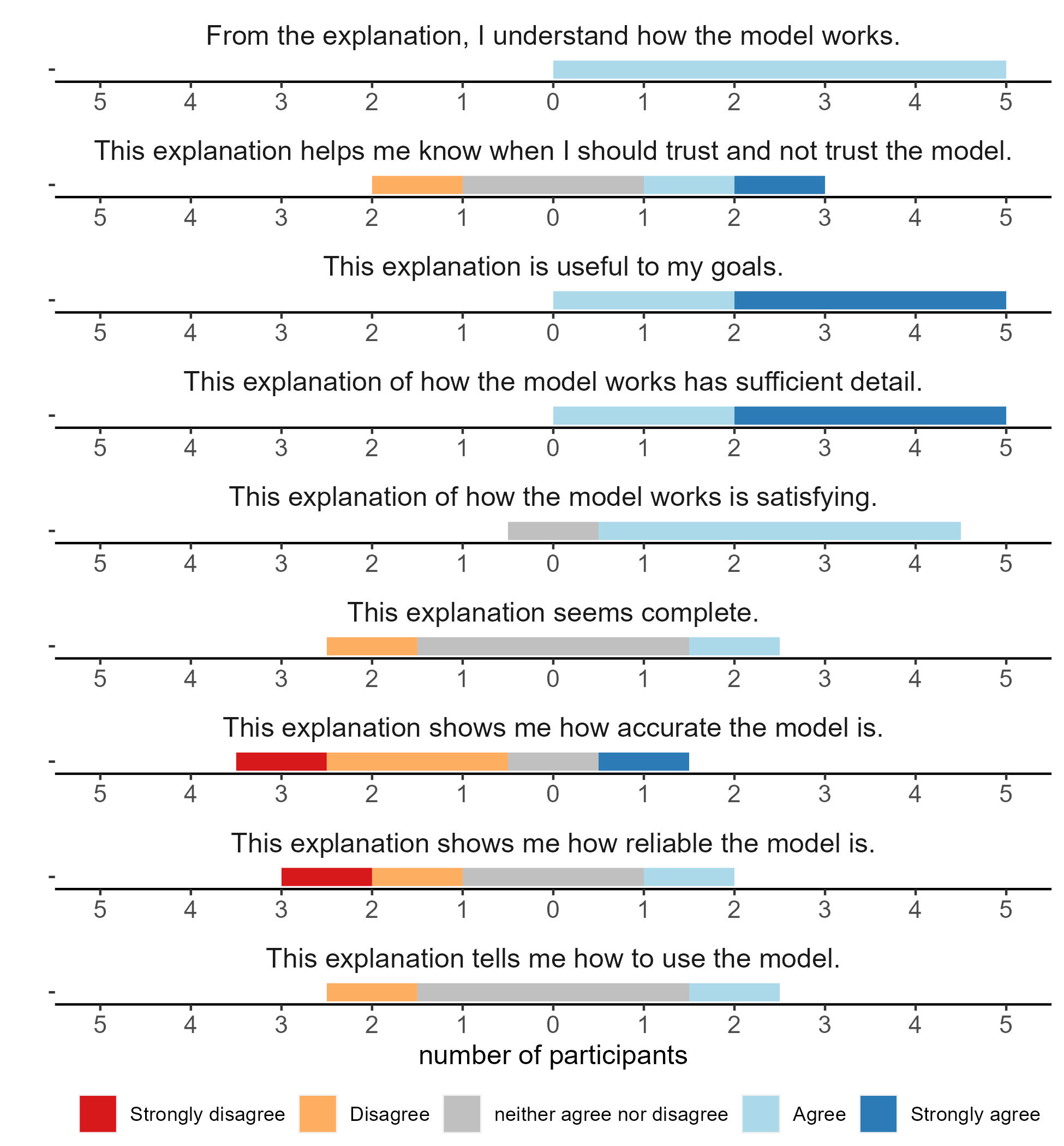}
    \caption{Answers to the explainability satisfaction scale. For each question, colored bars display the number of people per answer, centered around the neutral answer with disagreeing answers to the left and agreeing answers to the right.}
    \label{fig:ESG}
\end{figure}

The ESC gives an overview of how helpful our tool was seen (see Fig. ~\ref{fig:ESG}). Experts agreed that our explanation helps them understand the model, is satisfying, provides sufficient detail, and is useful for their goals. However, they were neutral regarding the completeness of the explanation and its guidance on using the model. Opinions varied on whether the explanation conveyed the model's accuracy, reliability, or trustworthiness.

One expert noted they had used a similar approach in their work, manually creating visualizations of different subsets of the data. 
Participants noted that the tool is limited to tabular data, restricting its applicability. Another pointed out that our fuzzy definition of neighborhoods might be problematic for users seeking specific values. 

\section{Discussion}
Our case studies highlight the effectiveness and generalizability of our appoach. With \textsc{FINCH}, we found a wide variety of feature interactions in multiple data sets, without being limited to only two dimensions. 
All the experts were motivated to use our tool and noted in the final questionnaire that it was useful for their goals. The only limitation mentioned was that the tool is designed exclusively for tabular data, while experts also need tools for other data types, such as images.

The core idea of using subsets was intuitive to all participants, but clearly specifying what is included in the current subset proved crucial. The tools' options for calibrating trust, like ground truth and uncertainty visualizations, were 
well-received and deemed important. 
A limitation of the ground truth view is the impact of mean approximation on the ground truth curve; limited ground truth values per feature may not accurately represent the true mean, affecting result interpretation. Nevertheless, this view offers valuable insight into whether discrepancies arise from the model or the data.
The most complex view was separating the main and interaction effects, which participants found challenging. Effectively using this view may require significant reworking or a more extensive tutorial. 

However, experts indicated that more global and quantitative data, such as R² scores, are needed to fully assess a model's trustworthiness.
The experts also shared several suggestions for improving the tool, mostly focusing on providing clearer descriptions and enhancing the tool's interactivity. We observed all experts frequently adjusting feature values to see how predictions changed, underscoring the necessity and significant potential for interactive explanations in this context.
 
Overall, the tool was well received, and expert feedback confirmed that it met our requirements. The only questionnaire with varied responses was the ESC, reflecting the tool's specific focus. While the tool was seen as useful, it was not intended or perceived as providing complete explanations or fully assessing the model’s trustworthiness; global explanations are more suitable for that purpose. Instead, the tool successfully achieved its stated goals, and feedback on those objectives was highly positive.

\section{Conclusion}
Current xAI approaches often focus on attributing relevance to individual features, overlooking their influential interactions. Our approach advances explainable AI by enabling the visualization of local feature interactions up to high orders. Our tool, \textsc{Finch}, employs a subset-based method to create intuitive visualizations of feature interactions relevant to a specific data instance, while also offering functionalities to validate displayed curves and help users calibrate their trust in the results. Our approach can be adapted to any machine learning model.
Our case studies demonstrated our tools' generalizability, and a user study with five machine learning experts confirmed \textsc{Finch}'s high usability. Using \textsc{Finch}, machine learning experts can examine how high-order feature interactions influence any data instance.

Future research includes generating more reliable results when data instances are sparse at the edges of distributions. Further work is also needed to automatically detect local feature interactions. Given the frequent use of other data types, such as images, future efforts could adapt this approach to those formats.
We also want to provide more support for analyzing feature interactions directly on the ground truth, instead of using the model.
Lastly, we aim to simplify our approach to make it more intuitively understandable for a broader audience, including domain experts and the general public.

\bibliographystyle{abbrv-doi-hyperref}
\bibliography{template.bib}

\begin{thebibliography}{10}

\bibitem{Inglis2022Visualizing}
A.~P. Alan~Inglis and C.~B. Hurley.
\newblock Visualizing variable importance and variable interaction effects in machine learning models.
\newblock {\em Journal of Computational and Graphical Statistics}, 31(3):766--778, 2022.

\bibitem{apley2020visualizing}
D.~W. Apley and J.~Zhu.
\newblock Visualizing the effects of predictor variables in black box supervised learning models.
\newblock {\em Journal of the Royal Statistical Society Series B: Statistical Methodology}, 82(4):1059--1086, 2020.

\bibitem{bertrand2022cognitive}
A.~Bertrand, R.~Belloum, J.~R. Eagan, and W.~Maxwell.
\newblock How cognitive biases affect xai-assisted decision-making: A systematic review.
\newblock In {\em Proceedings of the 2022 AAAI/ACM Conference on AI, Ethics, and Society}, pp. 78--91, 2022.

\bibitem{britton2019vine}
M.~Britton.
\newblock Vine: Visualizing statistical interactions in black box models.
\newblock {\em arXiv preprint arXiv:1904.00561}, 2019.

\bibitem{Brooke1996SUS}
J.~Brooke.
\newblock Sus -- a quick and dirty usability scale.
\newblock In {\em Usability Evaluation in Industry}, pp. 189--194, 01 1996.

\bibitem{BRFSS2015}
{Centers for Disease Control and Prevention (CDC)}.
\newblock Behavioral risk factor surveillance system survey data.
\newblock U.S. Department of Health and Human Services, Centers for Disease Control and Prevention, 2015.

\bibitem{danilevsky2020survey}
M.~Danilevsky, K.~Qian, R.~Aharonov, Y.~Katsis, B.~Kawas, and P.~Sen.
\newblock A survey of the state of explainable ai for natural language processing.
\newblock {\em arXiv preprint arXiv:2010.00711}, 2020.

\bibitem{dawson1995unusual}
R.~J.~M. Dawson.
\newblock The “unusual episode” data revisited.
\newblock {\em Journal of Statistics Education}, 3(3), 1995.

\bibitem{mijolla2020human}
D.~de~Mijolla, C.~Frye, M.~Kunesch, J.~Mansir, and I.~Feige.
\newblock Human-interpretable model explainability on high-dimensional data.
\newblock {\em arXiv preprint arXiv:2010.07384}, 2021.

\bibitem{eiband2019impact}
M.~Eiband, D.~Buschek, A.~Kremer, and H.~Hussmann.
\newblock The impact of placebic explanations on trust in intelligent systems.
\newblock In {\em CHI Conference on Human Factors in Computing Systems Extended Abstracts}, pp. 1--6, 2019.

\bibitem{fanaee2014event}
H.~Fanaee-T and J.~Gama.
\newblock Event labeling combining ensemble detectors and background knowledge.
\newblock {\em Progress in Artificial Intelligence}, 2:113--127, 2014.

\bibitem{ferrettini2022coalitional}
G.~Ferrettini, E.~Escriva, J.~Aligon, J.-B. Excoffier, and C.~Soul{\'e}-Dupuy.
\newblock Coalitional strategies for efficient individual prediction explanation.
\newblock {\em Information Systems Frontiers}, 24(1):49--75, 2022.

\bibitem{figueira2022survey}
A.~Figueira and B.~Vaz.
\newblock Survey on synthetic data generation, evaluation methods and gans.
\newblock {\em Mathematics}, 10(15):2733, 2022.

\bibitem{friedman2001greedy}
J.~H. Friedman.
\newblock Greedy function approximation: a gradient boosting machine.
\newblock {\em Annals of statistics}, pp. 1189--1232, 2001.

\bibitem{friedman2024function}
J.~H. Friedman.
\newblock Function trees: Transparent machine learning.
\newblock {\em arXiv preprint arXiv:2403.13141}, 2024.

\bibitem{herbinger2022Repid}
J.~Herbinger, B.~Bischl, and G.~Casalicchio.
\newblock {REPID: R}egional effect plots with implicit interaction detection.
\newblock In {\em International Conference on Artificial Intelligence and Statistics}, pp. 10209--10233. PMLR, 2022.

\bibitem{herbinger2023decomposing}
J.~Herbinger, M.~N. Wright, T.~Nagler, B.~Bischl, and G.~Casalicchio.
\newblock Decomposing global feature effects based on feature interactions.
\newblock {\em arXiv preprint arXiv:2306.00541}, 2023.

\bibitem{hoffman2018metrics}
R.~R. Hoffman, S.~T. Mueller, G.~Klein, and J.~Litman.
\newblock Metrics for explainable ai: Challenges and prospects.
\newblock {\em arXiv preprint arXiv:1812.04608}, 2018.

\bibitem{hohman2019gamut}
F.~Hohman, A.~Head, R.~Caruana, R.~DeLine, and S.~M. Drucker.
\newblock Gamut: A design probe to understand how data scientists understand machine learning models.
\newblock In {\em Proceedings of the 2019 CHI conference on human factors in computing systems}, pp. 1--13, 2019.

\bibitem{holzinger2022explainable}
A.~Holzinger, A.~Saranti, C.~Molnar, P.~Biecek, and W.~Samek.
\newblock Explainable ai methods-a brief overview.
\newblock In {\em xxAI - Beyond Explainable AI: International Workshop}, pp. 13--38, 2022.

\bibitem{jullum2021efficient}
M.~Jullum, A.~A. Redelmeier, and K.~Aas.
\newblock Efficient and simple prediction explanations with groupshapley: a practical perspective.
\newblock In {\em Italian Workshop on Explainable Artificial Intelligence}, vol. 3014, 2021.

\bibitem{kaur2020interpreting}
H.~Kaur, H.~Nori, S.~Jenkins, R.~Caruana, H.~Wallach, and J.~Wortman~Vaughan.
\newblock Interpreting interpretability: understanding data scientists' use of interpretability tools for machine learning.
\newblock In {\em Proceedings of the 2020 CHI conference on human factors in computing systems}, pp. 1--14, 2020.

\bibitem{krause2016interacting}
J.~Krause, A.~Perer, and K.~Ng.
\newblock Interacting with predictions: Visual inspection of black-box machine learning models.
\newblock In {\em Proceedings of the 2016 CHI conference on human factors in computing systems}, pp. 5686--5697, 2016.

\bibitem{lundberg2018consistent}
S.~M. Lundberg, G.~G. Erion, and S.-I. Lee.
\newblock Consistent individualized feature attribution for tree ensembles.
\newblock {\em arXiv preprint arXiv:1802.03888}, 2018.

\bibitem{miller2019explanation}
T.~Miller.
\newblock Explanation in artificial intelligence: Insights from the social sciences.
\newblock {\em Artificial intelligence}, 267:1--38, 2019.

\bibitem{molnar2023model}
C.~Molnar, G.~K{\"o}nig, B.~Bischl, and G.~Casalicchio.
\newblock Model-agnostic feature importance and effects with dependent features: a conditional subgroup approach.
\newblock {\em Data Mining and Knowledge Discovery}, pp. 1--39, 2023.

\bibitem{pace1997sparse}
R.~K. Pace and R.~Barry.
\newblock Sparse spatial autoregressions.
\newblock {\em Statistics \& Probability Letters}, 33(3):291--297, 1997.

\bibitem{ribeiro2016should}
M.~T. Ribeiro, S.~Singh, and C.~Guestrin.
\newblock " why should i trust you?" explaining the predictions of any classifier.
\newblock In {\em Proceedings of the 22nd ACM SIGKDD international conference on knowledge discovery and data mining}, pp. 1135--1144, 2016.

\bibitem{seedorff2021totalvis}
N.~Seedorff and G.~Brown.
\newblock totalvis: A principal components approach to visualizing total effects in black box models.
\newblock {\em SN Computer Science}, 2(3):141, 2021.

\bibitem{tsang2021interpretable}
M.~Tsang, J.~Enouen, and Y.~Liu.
\newblock Interpretable artificial intelligence through the lens of feature interaction.
\newblock {\em arXiv preprint arXiv:2103.03103}, 2021.

\bibitem{Wang2019DesigningTU}
D.~Wang, Q.~Yang, A.~Abdul, and B.~Y. Lim.
\newblock Designing theory-driven user-centric explainable ai.
\newblock {\em Proceedings of the 2019 CHI Conference on Human Factors in Computing Systems}, 2019.

\bibitem{xin2024you}
X.~Xin, F.~Huang, and G.~Hooker.
\newblock Why you should not trust interpretations in machine learning: Adversarial attacks on partial dependence plots.
\newblock {\em arXiv preprint arXiv:2404.18702}, 2024.

\bibitem{zhang2023capturing}
H.~Zhang, X.~Zhang, T.~Zhang, and J.~Zhu.
\newblock Capturing the form of feature interactions in black-box models.
\newblock {\em Information Processing \& Management}, 60(4):103373, 2023.

\end{thebibliography}

\section*{Suppplementary Material}
The code is open source and available on GitHub\footnote{https://github.com/akleinau/Finch}. Any currently running online versions of the tool can be found there too.

\appendix

\section*{Appendix A}
\label{AppendixA}

We used a series of data sets in our case studies and as examples in our paper. 

\subsection*{Bike Sharing}
The bike sharing data set is used to predict the number of bike rentals per hour. 

We trained a MLPC Regression model.

We used an 80:20 train:test split resulting in 13903 instances being used for training and in \textsc{Finch}.

In our example case, we used the following features:
\begin{itemize}
    \item count (target): the number of bike rentals that hour
    \item hour=3: the hour for which the bike rentals where recorded. Here: 3am.
    \item workingday=0: if the instance was recorded on a workingday or not. Categoric feature. 0=no, 1=yes.
    \item season=0: in which season the instance was recorded. Categoric feature. 0=winter, 1=spring, 2=summer, 3=autumn.
\end{itemize}

\begin{figure}[h!]
    \centering
    \begin{subfigure}[b]{0.5\textwidth}
        \centering
        \includegraphics[width=1\linewidth]{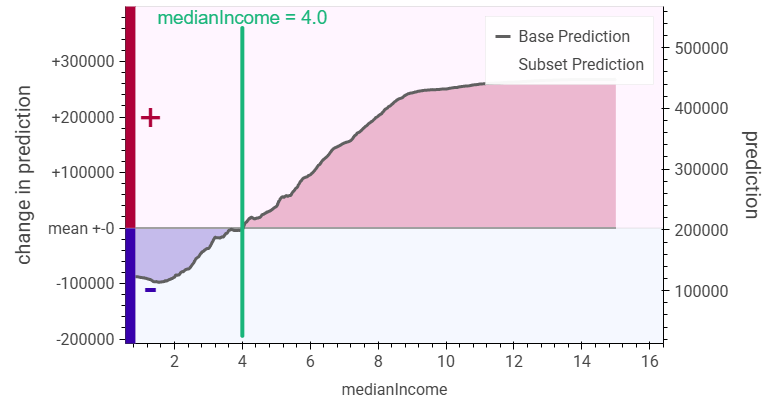}
        \caption{house value per median income}
        \label{fig:california1}
    \end{subfigure}
    \hspace{0.001\textwidth} % Adds some space between the two images
    \begin{subfigure}[b]{0.5\textwidth}
        \centering
        \includegraphics[width=1\linewidth]{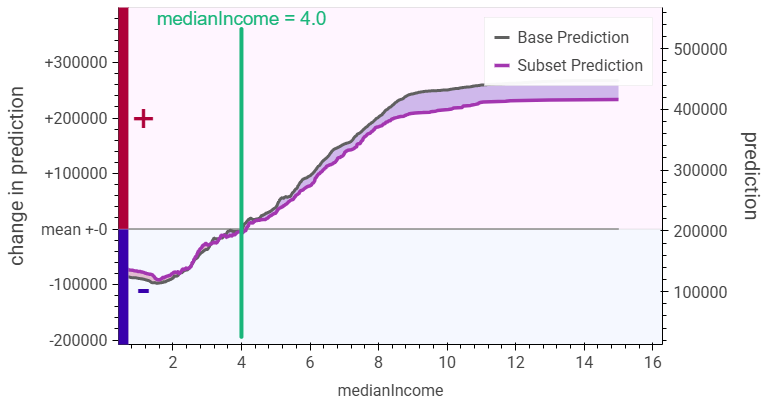}
        \caption{house value per median income for areas with a population of 2000}
        \label{fig:california2}
    \end{subfigure}
    \hspace{0.001\textwidth} % Adds some space between the two images
    \begin{subfigure}[b]{0.5\textwidth}
        \centering
        \includegraphics[width=1\linewidth]{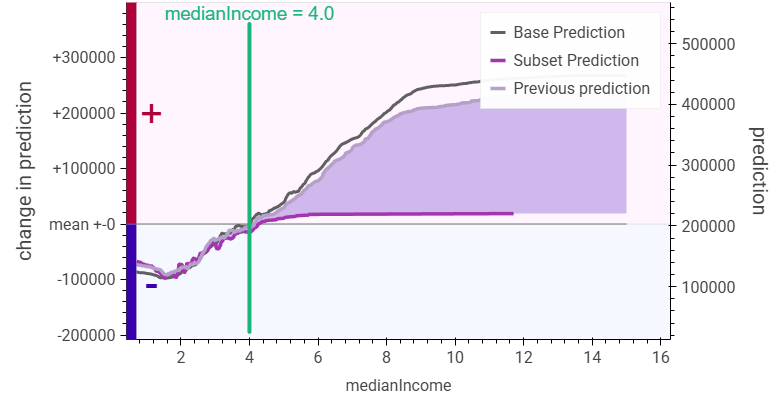}
        \caption{house value per median income for areas with a population of 2000 and 1000 total rooms}
        \label{fig:california3}
    \end{subfigure}
    \hspace{0.001\textwidth} % Adds some space between the two images
    \begin{subfigure}[b]{0.5\textwidth}
        \centering
        \includegraphics[width=1\linewidth]{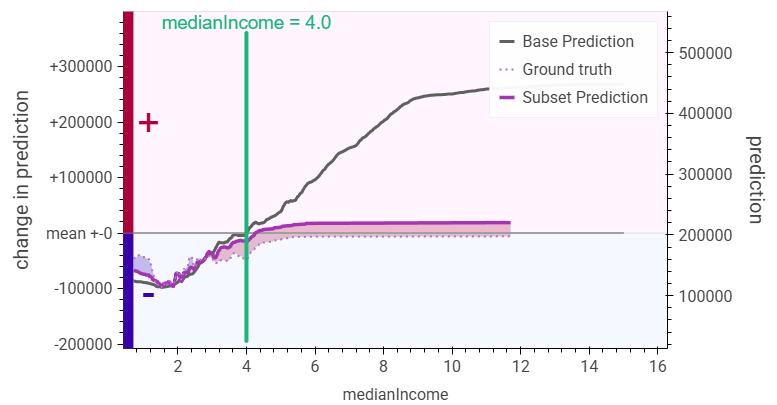}
        \caption{The ground truth is even lower.}
        \label{fig:california4}
    \end{subfigure}
    \caption{The incremental visualization of the interaction of median income, population and total rooms in the california housing data set. Colored areas visualize the change in each step.}
\end{figure}

\subsection*{Titanic}
The titanic data set is used to predict the survival of people on board the titanic. 

We trained an MLPC classifier. The resulting accuracy was 70.23\%.

We used an 80:20 train:test split resulting in 1047 instances being used for training and in \textsc{Finch}.

In the described interaction, we used the following features:
\begin{itemize}
    \item survival(target): If the current person survived.
    \item pclass=1: Which passenger class the current person belonged to. Categoric feature. 1=first class, 2=second class, 3=third class. 
    \item sex=1: The sex of the person. Categoric feature. 0=male, 1=female.
    \item age:30: The age of the person. Here: 30yo.
\end{itemize}

\subsection*{California housing}
The california housing data set is used to predict housing values for block groups in California and was derived from the 190 US census. It contains only continuous variables. The mean predicted housing value is 200.000. 

We trained a GradientBoostingRegressor model on the data. 
It was trained with 100 boosting stages, a learning rate of 0.1 and squared error as the loss function.
The resulting R2 score was 0.77.

We used an 80:20 train:test split resulting in 16,512 instances being used for training and in \textsc{Finch}.

In our observed interaction, we used the following features:
\begin{itemize}
    \item housing value (target): Median house value in US Dollars.
    \item median income: The median income of that block group in 100,000 US Dollars.
    \item population: The number of people residing in the block group.
    \item total rooms: The total number of rooms in that block group.
\end{itemize}

\subsection*{Diabetes}
The diabetes risk factor data set. It is based on the BRFSS telephone study that is performed yearly in the united states.

We used a subset of 10,000 instances for model training. Using a 80:20 train/test split, this resulted in 8000 instnaces being used in \textsc{Finch}.

For better model training, half of the instances are diabetes positive, and half negative. Therefore, the probabilities generated by the model and \textsc{finch} cannot be directly used on a general public.

In our interaction, we considered the following features:
\begin{itemize}
    \item diabetes risk (target): The diabetes risk for the person, that the model predicted.
    \item sex=0: The sex of the person. 0=male, 1=female.
    \item exercise=1: If the person exercises. 1=yes, 0=no.
    \item high blood pressure=1: If the person has high blood pressure. Categoric feature. 1=yes, 0=no.
    \item weight category=3: The weight category of the person. Categoric feature. 0=underweight, 1= normal weight, 2=overweight, 3=obese.  
\end{itemize}

\begin{figure}[h]
    \centering
    \begin{subfigure}[b]{0.5\textwidth}

        \includegraphics[width=0.7\linewidth]{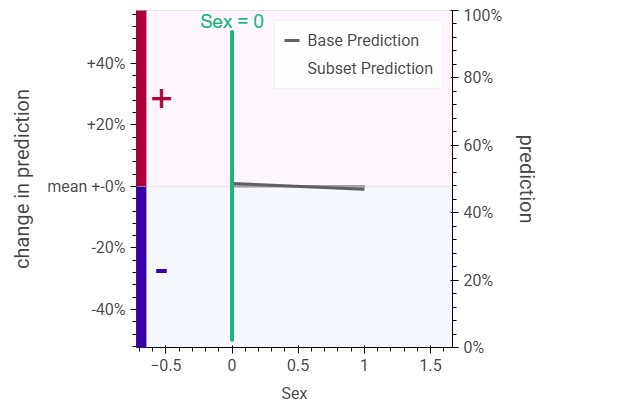}
        \caption{diabetes risk per sex}
        \label{fig:diabetes1}
    \end{subfigure}
    \hspace{0.001\textwidth} % Adds some space between the two images
    \begin{subfigure}[b]{0.5\textwidth}
        \centering
        \includegraphics[width=1\linewidth]{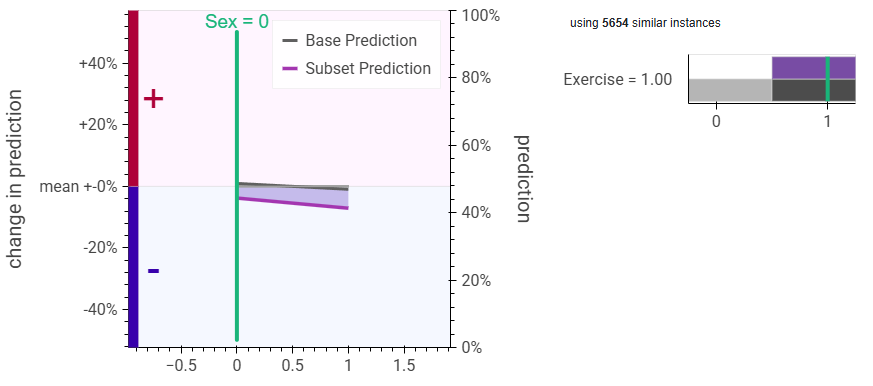}
        \caption{diabetes risk per sex for people who exercise}
        \label{fig:diabetes2}
    \end{subfigure}
    \hspace{0.001\textwidth} % Adds some space between the two images
    \begin{subfigure}[b]{0.5\textwidth}
        \centering
        \includegraphics[width=1\linewidth]{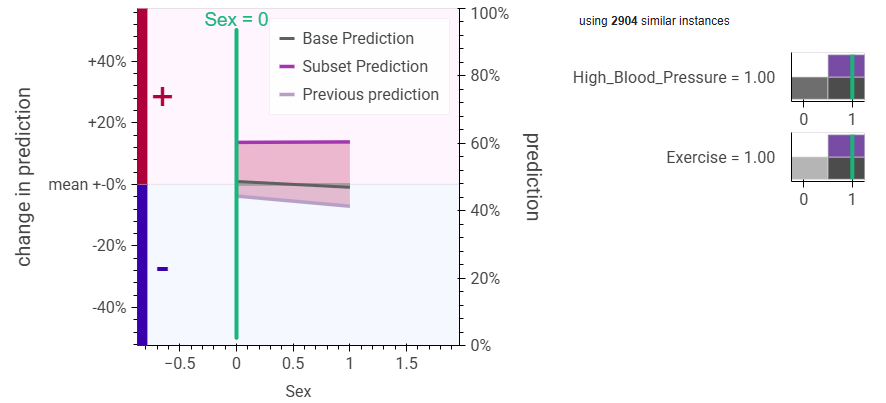}
        \caption{diabetes risk per sex for people who exercise but have high blood pressure}
        \label{fig:diabetes3}
    \end{subfigure}
    \hspace{0.001\textwidth} % Adds some space between the two images
    \begin{subfigure}[b]{0.5\textwidth}
        \centering
        \includegraphics[width=1\linewidth]{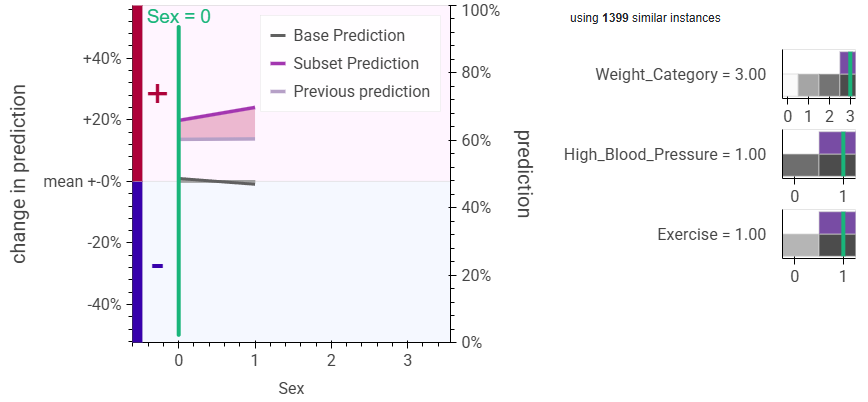}
        \caption{diabetes risk per sex for people who exercise, have high blood pressure and are obese}
        \label{fig:diabetes4}
    \end{subfigure}
    \caption{Diabetes risk per sex. Incremental interaction of sex, exercise, high blood pressure and weight. Based on the BRFSS data set.}
\end{figure}

\end{document}